\documentclass[acmtog, screen, noacm]{acmart}
\setcopyright{none}

\renewcommand\footnotetextcopyrightpermission[1]{}
\AtBeginDocument{%
  }

\usepackage{wrapfig}
\usepackage{algorithm}
\usepackage{algpseudocode}
\usepackage[dvipsnames]{xcolor}

\definecolor{dgreen}{rgb}{0.1, 0.5, 0.1}
\definecolor{dred}{rgb}{0.7, 0.0, 0.0}
\definecolor{dblue}{rgb}{0.1, 0.1, 0.5}

\DeclareMathOperator*{\argmin}{argmin} 

\algdef{SE}[DOWHILE]{Do}{doWhile}{\algorithmicdo}[1]{\algorithmicwhile\ #1}%

\begin{document}

\title{Variational r-Adaptive Cloth Simulation}

\author{Jiahao Wen}
\affiliation{
\institution{Adobe}
\country{USA}}
\affiliation{
 \institution{University of Southern California}
 \country{USA}} 
\email{jiahaow@usc.edu}

\author{Zhen Chen}
\affiliation{
\institution{Adobe}
\country{USA}}
\email{zhenc@adobe.com}

\author{Jernej Barbi\v{c}}
\affiliation{
\institution{University of Southern California}
\country{USA}}
\email{jnb@usc.edu}

\author{Danny M. Kaufman}
\affiliation{
\institution{Adobe}
\country{USA}}
\email{dannykaufman@gmail.com}

\begin{abstract}

We present the first r-adaptive method for simulating frictionally contacting cloth dynamics and statics within modern cloth simulation pipelines. Cloth is a thin, highly deformable structure whose visually salient behaviors (wrinkling, folding, buckling, and sharp contact features) require high effective spatial resolution. While r-adaptivity has been widely studied for volumetric solids, directly applying existing r-adaptive strategies to thin shells with piecewise-linear elements exposes two fundamental failure modes. First, under discretization, IP-driven r-adaptive optimization can be trapped in local minima with lesser-quality solutions and so suboptimal physical configurations. Second, the variational remesher can achieve artificially low incremental potential (IP) energy by collapsing elements rather than improving the physical solution: strongly ill-shaped elements invalidate the finite element (FE) space approximation that the IP objective relies on, making IP itself unreliable as a measure of solution quality. We identify these as two distinct but coupled barriers to variational r-adaptivity in thin structures.

We address both with a single \emph{degeneracy-activated quality regularization} that is inactive in regions of well-shaped elements, leaving the variational remesher free to generate local anisotropies and densification, and strongly active  only as elements approach degeneracy. This formulation suppresses spurious low-energy basins sourced from poor discretizations and reshapes the IP landscape to enable escape from suboptimal physical minima. In turn, this also enables an unbiased optimization of the physical objective in well-shaped regions and also resolves a cloth-specific failure mode we term \emph{element bunching}, where mesh elements progressively collapse as cloth slides over sharp contact features. To make r-adaptive cloth practical, we further introduce a dynamic nonlinear solver that exploits within-timestep coherence for r-adaptive ITR solves in derivative evaluation and dynamic IPC tolerance updating. This yields a \textbf{3--6X speedup} over prior optimal ITR~\cite{Wen:2025:ORI}. We demonstrate our method on challenging frictional contact scenarios, showing that r-adaptivity produces higher visual fidelity than fixed meshes under equal vertex-count and time-budget constraints.

\end{abstract}

\begin{teaserfigure}
  \centering
  \includegraphics[width=1.0\textwidth]{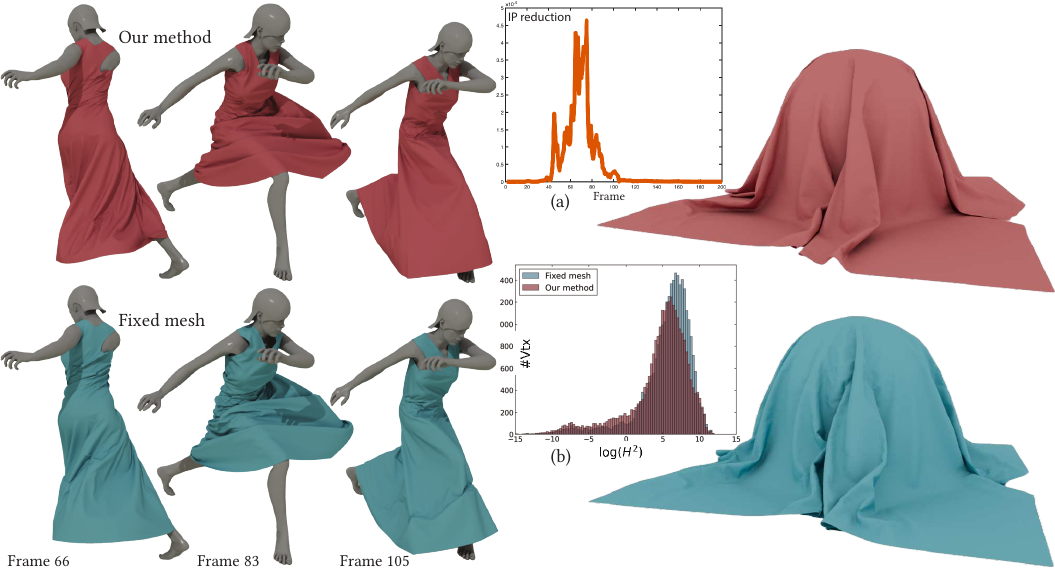}
  \vspace{-8mm}
  \caption{\textbf{High-fidelity cloth simulation with r-adaptivity.} We demonstrate our method on both dynamic and static contact scenarios. \textbf{Left (dynamic):} we simulate a dress driven by a mannequin performing extremely large and rapid motions (frames 66, 83, 105 shown; \#V: 14K). The fixed mesh simulation (bottom) suffers from severe membrane locking and produces aliased wrinkle patterns, while our method resolves smooth wrinkle flows (Frame 66), preserves complex multi-layered folds (Frame 83), and produces natural draping as the motion settles (Frame 105) without long locked creases along the bottom. \textbf{Right (static):} a square cloth drapes over a sphere using identical vertex counts (\#V: 31K). The fixed mesh simulation (bottom) produces blocky, membrane-locked folds with the cloth unnaturally folded along both the sphere and ground, while our method (top) produces smooth, continuous folds that conform naturally to the sphere and ground geometries. \textbf{Middle (quantitative comparison):} In (a) we plot per-frame IP reduction $E_{\text{fixed}}-E_{\text{ours}}$ of our method over the dynamic animation sequence, with peak reduction coinciding with the most intense deformation phases (frames 40--100) of the dancing; and (b) per-vertex distribution of $\log(H^2)$ (squared mean curvature) for the sphere example, where the fixed mesh exhibits more high-curvature concentration in the right tail while our method shifts the distribution toward lower curvature.}
  \label{fig:teaser}
\end{teaserfigure}

\maketitle

\section{Introduction}
Cloth simulation is essential to computer graphics, VFX, games, and fashion. Simulated with thin-shell, high-quality cloth modeling is generally characterized by capturing detailed folding, buckling and conforming contact. In turn, however, this generally requires high-resolution meshes. To avoid working with expensive, uniformly high-resolution meshes, adaptive simulation methods work to apply resolution where it can be most effective.
Generally, there are three primary strategies for mesh-adaptivity in simulation: h-adaptivity (inserting additional vertices), p-adaptivity (increasing the order of interpolation functions), and r-adaptivity (moving the vertices of the undeformed, ``reference'' mesh to spatially increase precision wherever necessary). In this paper, we focus on r-adaptivity for cloth, and give the first r-adaptive cloth method that works with modern high-fidelity shell simulation pipelines using the Incremental Potential Contact (IPC) model~\cite{Li:2020:IPC,Li2021CIPC}.

Although r-adaptivity has previously been investigated for volumetric solids, cloth presents unique challenges. Cloth is highly bendable and requires higher mesh resolutions to capture visually salient features such as folds and wrinkles. As we show, applying state-of-the-art volumetric r-adaptive strategies directly to cloth simulation does not work as there exist fundamental failure modes that must be addressed. We begin from the variational formulation of r-adaptive dynamic simulation via In-Timestep Remeshing (ITR)~\cite{Ferguson:2023:ITR,Wen:2025:ORI}, which poses r-adaptivity as a per-timestep optimization over the full simulation mesh. This formulation provides an attractive remeshing objective that integrates all system energies when choosing a best-fit reference mesh. 

However, when applied to thin shells with low-order (here linear) elements, this objective admits a previously unaddressed pathology: \textit{the variational remesher can achieve artificially low incremental potential (IP) by collapsing elements rather than improving the physical solution.} The root cause is that strongly ill-shaped elements generate large discretization errors~\cite{Shewchuk:2002:WIA}, locally invalidating the FE-space approximation that the IP objective relies on. This issue is especially problematic and unavoidable for the low-order elements and thin-shell bending models employed in graphics. Once the FE approximation degrades, the IP energy itself ceases to be a reliable measure of physical solution quality. 
Compounding this issue, IP-driven simulation is inherently nonconvex, with contact and large-deformation elasticity giving rise to many local minima. In the spatially continuous setting all minima are equally valid physical solutions enabled by the underlying nonconvexity of the elastic energy. However, \emph{under discretization}, we observe that minima of the IP are no longer all of equivalent quality: IP-based r-adaptivity can then be trapped in basins supporting lesser-quality physical configurations. We identify these as two distinct but coupled barriers to variational r-adaptivity in thin structures, neither resolvable by the post-hoc mesh smoothing strategies common in prior works.

Our first contribution is a \emph{degeneracy-activated quality regularization} that resolves both pathologies at the same time. The regularizer is constructed to be inactive in regions of well-shaped elements, leaving the variational remesher free to generate appropriately aggressive local mesh anisotropies and densification,  and only activates as elements approach severely distorted mesh configurations or degeneracies. This unified formulation simultaneously suppresses spurious low-energy basins from bad mesh quality and reshapes the IP landscape to enable the solver to escape poor-quality solution minima, all while preserving the unbiased optimization of the underlying physical objective in well-shaped regions. It also resolves a specific failure that we term \textit{element bunching}: when cloth slides across a sharp edge of a colliding object (e.g., a table edge), the elements on one side of the edge progressively become entangled or collapsed under naive r-adaptive remeshing, causing simulation artifacts (Figure~\ref{fig:bunching}). Our regularization naturally prevents this problem, enabling crisp and stable resolution of contact features that have driven the appeal of r-adaptivity in contact-rich scenarios from the start.

Our second contribution addresses performance. R-adaptive ITR methods are expensive~\cite{Wen:2025:ORI}, and cloth simulation requires significant degree-of-freedom (DOF) counts to resolve fine wrinkles and folds. We accelerate ITR-based r-adaptive cloth simulation through a new solver that exploits temporal continuity structure in the ITR optimization process for efficient derivative updates, combined with dynamic tolerance adjustments via inexact Newton iterations. Together these improvements yield a \textbf{3--6X speedup} over previous state-of-the-art ITR solves~\cite{Wen:2025:ORI}, making higher-fidelity r-adaptive cloth simulation practical.

We demonstrate our method on challenging frictionally contacting cloth scenarios, showing that our r-adaptive simulations outperform fixed-mesh baselines both quantitatively and qualitatively at equal vertex budget, while remaining advantageous under \emph{equal computational time budget}: yielding cleaner silhouettes, less tangling, and smoother, better-resolved folds. Ablation studies isolate each algorithmic contribution and confirm clear degradation when prior volumetric ITR methods are applied naively to cloth.

\section{Related Work}
\subsection{Thin-Shell Cloth Simulation}
Thin-shell mechanics has long been central to both computational mechanics and computer graphics, particularly for cloth simulation~\cite{Terzopoulos1987,Bridson2002,Grinspun:2003:DS,Goldenthal2007,Narain:2012:AAR,Narain:2013:FAC,LI:2018:AIF}. Substantial progress spans implicit time-integration~\cite{Baraff1998,Bridson2002,Kim2020}, contact and collision handling~\cite{Tang2016CAMA,Tang2018ICloth,Li2021CIPC,Ando2024CBD,Huang2024:GIPC}, parallelism and high-performance implementations~\cite{Macklin2016XPBD,Chen2024VBD}, expressive constitutive models~\cite{Weischedel:2012:ADG,Chen:2018:PSO,Chen2021TFC,Wen:2023:KLS}, and adaptive discretization for robustness under large deformations~\cite{Grinspun:2002:CAS,Narain:2012:AAR,Narain:2013:FAC,Li:2005:CAW,Pfaff:2014:ATA,Bender:2013:ACS,Villard:2005:AMF,Simnett:2009:AEA}.
\subsection{Adaptive Discretization in Cloth Simulation}
High-fidelity cloth simulation requires fine discretization to resolve localized features and complex deformation, but uniform refinement is computationally prohibitive under implicit integration. Adaptive strategies dynamically redistribute degrees of freedom to balance accuracy and efficiency~\cite{Manteaux:2017:APB}.
\paragraph{h- and p-adaptive methods.}
Most adaptive cloth methods fall into h-adaptivity, locally refining or coarsening the mesh through vertex insertion, deletion, or topology modification, typically guided by geometric criteria such as strain, curvature, or bending energy~\cite{Vasilescu:1992:AMA,Li:2005:CAW,Narain:2012:AAR,Narain:2013:FAC,LI:2018:AIF}. These methods are largely driven by local measures without a global optimization objective, leading to suboptimal refinement patterns and convergence to poor local minima. Basis refinement (p-adaptivity) has also been applied~\cite{Grinspun:2002:CAS} with similar challenges in formulating refinement measures.
\paragraph{r-adaptive methods.}
An alternative is r-adaptivity, where mesh connectivity remains fixed but vertex positions are dynamically adjusted. \citet{Cho:2004:RAM} propose an r-adaptive method redistributing vertices based on localized stress analysis on high-order elements. While effective for some equilibrium stress evaluations, this approach handles neither dynamics nor contact — both essential for cloth simulation.
\paragraph{Eulerian-on-Lagrangian methods.}
A closely related family is Eulerian-on-Lagrangian (EoL) simulation~\cite{Weidner:2018:ECS, Wen:2020:CRW}, which decouples material and spatial coordinates to allow vertices to slide over collider geometry. EoL relies on heuristic h-adaptivity to insert vertices at contact points, includes referential coordinate velocities (introducing nullspace ambiguities resolved through artificial constraints), and requires explicit treatment for conforming contacts. In contrast, our ITR-based framework does not model referential coordinate dynamics; coupling between deformed positions and reference coordinates is preserved in-solve through physical admissibility, and conforming contacts are handled implicitly through the variational remeshing objective. Our method preserves the underlying problem symmetries (Figure~\ref{fig:spikes}) and avoids EoL's heuristic interventions.
\subsection{Physics-Guided Adaptive Simulation}
Beyond cloth, physics-guided adaptive discretization has been extensively studied in volumetric elastodynamics. Early works guide adaptivity through elastic energy or continuum-mechanics error estimates~\cite{Demkowicz:2006:CWH,Mitchell:2014:ACO}. \citet{Mosler:2006:OTN,Mosler:2007:VHI} introduce optimization of incremental potential energy as a principled objective for mesh adaptivity in contact-free elastostatics and plasticity. \citet{Ferguson:2023:ITR} extend this to the contacting dynamics in-timestep remeshing (ITR) framework, performing h-adaptivity within each timestep, and \citet{Wen:2025:ORI} extend ITR to r-adaptive settings; see~\cite{Wen:2025:ORI} for a detailed discussion. To our knowledge, no prior work applies physics-guided r-adaptive techniques to thin-shell cloth, motivating our approach.

\section{Formulation}\label{sec:itr_framework}
We build our r-adaptive thin-shell framework upon the variational In-Timestep Remeshing (ITR) formulation~\cite{Ferguson:2023:ITR,Wen:2025:ORI}, extended to shell elastica with several shell-specific components. We first review the thin-shell elasticity model used throughout this work, then describe our ITR formulation for shells.

\subsection{Thin-Shell Background}
We model thin shells using the Kirchhoff-Love assumption, where the volumetric deformation is fully determined by the midsurface deformation $x(X,t):\Omega \times \mathbb{R}^+ \rightarrow \mathbb{R}^3$ over a UV parameter domain $\Omega \subset \mathbb{R}^2$. We focus on rest-flat shells (plates) and identify $\Omega$ with the rest configuration; extension to rest-curved shells is left to future work. For the elastic energy density, we adopt the volumetric-reduced shell formulation\footnote{Alternately, popular and efficient hinge-based models like Discrete Shells~\cite{Grinspun:2003:DS} are similarly applicable here. Since it's orthogonal to our method, we leave the exploration to future work.} of~\cite{Chen:2018:PSO,Weischedel:2012:ADG,Wen:2023:KLS}, which yields the areal density via unified treatment of in-plane stretching and out-of-plane bending contributions. We refer readers to~\cite{Wen:2023:KLS} for complete kinematic formulation. Energy derivatives w.r.t. both world coordinates $x$ and reference coordinates $u$ are in the supplemental.

\subsection{ITR Framework for Shells}

\paragraph{Semi-discrete incremental potential.} Following~\citet{Ferguson:2023:ITR}, we discretize in time while staying spatially continuous, posing each implicit timestep as the optimization
\begin{equation}
x^{t+1} = \arg\min_x E_t(x),
\end{equation}
where $E_t$ is the spatially continuous IP integrating inertial, elastic, contact, and friction contributions over $\Omega$ (see~\cite{Ferguson:2023:ITR,Li:2020:IPC} for the explicit form). For demonstration we use implicit Euler timestepping; alternative integrators apply directly. We adopt the projection-corrected inertial model of~\citet{Wen:2025:ORI} for optimal ITR consistency.

\paragraph{Spatial discretization.} We discretize the midsurface using a triangle mesh $\mathcal{T}=(u,e)$ with $n$ nodal reference coordinates $u_i \in \Omega$ stored in $u=(u_1^T,...,u_n^T)^T \in \mathbb{R}^{2n}$, and connectivity $e$. Midsurface fields are stored at vertices as $x,v \in \mathbb{R}^{3n}$ (positions and velocities). Each term in the IP is approximated as a weighted sum over local element stencils, $\Sigma_{s\in \mathcal{T}}w_s(u)W_s(x,u)$, where $w_s$ is the rest-area-weighted scaling and $W_s$ is the per-element energy density.

\paragraph{Optimal ITR formulation.} We variationally adapt the reference mesh $u$ jointly with the deforming positions $x$. At each timestep, following~\citet{Wen:2025:ORI}, we seek a new mesh $u^{t+1}$ and its corresponding configuration $x^{t+1}$ that locally minimize a remeshing objective $R(x,u)$, subject to geometric admissibility $u^{t+1}\in\mathcal{G}$, (i.e. injective and shape-preserving~\cite{Wen:2025:ORI}) and the physical timestep update rule. We solve this as a nested bi-level optimization,
\begin{gather} \label{eq:optimization}
    \begin{split}
        u^{t+1} = &\argmin_u W(u), \> \text{s.t.} \> u \in \mathcal{G} \> \textrm{with}\\
        &W(u) = R(\argmin_{x} E_t(x,u),u).
    \end{split}
\end{gather}

\paragraph{Remeshing objective and shell-specific challenges.} Following~\citet{Ferguson:2023:ITR} and~\citet{Wen:2025:ORI}, we use the IP as the remeshing objective augmented with an optional regularization term:
\begin{equation}
R(x, u) = E_t(x, u) + C(u).
\end{equation}
While prior ITR works treat $C(u)$ as a nice-to-have augmentation, we identify a previously unrecognized necessity: for shell ITR, $C(u)$ is indispensable for resolving fundamental pathologies of the variational formulation, as analyzed in Section~\ref{sec:quality}. In addition, applying the ITR framework to shells introduces two practical challenges. First, the IP includes a local $L^2$ projection operator $\pi_t$ mapping inertial quantities from $u^t$ to each candidate mesh $u'$, requiring expensive derivatives of $\pi_t$ via chain rule~\cite{Leger:2014:AUL,Wen:2025:ORI} — a fundamental performance bottleneck for high-DOF shell simulations. We address this in Section~\ref{sec:acceleration}. Second, shell elasticity requires non-standard derivatives with respect to reference coordinates $u$; we provide analytic, efficient expressions for these in the supplemental. These shell-specific contributions, together with our solver acceleration strategies, enable robust and practical r-adaptive cloth simulation in challenging contact-rich settings.

\begin{figure}[!b]
    \centering
    \includegraphics[width=1\linewidth]{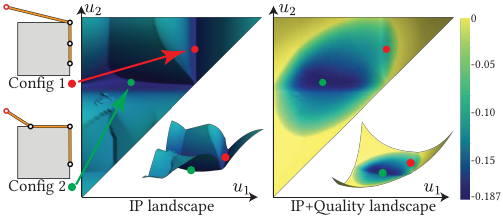}
    \vspace{-8mm}    
    \caption{\textbf{Quality regularization reshapes the IP landscape.} Consider a one-dimensional rod with four vertices: the first is pinned, the endpoints have fixed reference coordinates $u_0=0$ and $u_3=1$, and the rod contacts a rigid square under gravity. Enumerating $(u_1,u_2)$ reveals two non-equivalent local minima: a suboptimal minimum (red, configuration 1) supported by an over-stretched segment, and a global minimum (green, configuration 2) with well-shaped elements. Left: the original IP landscape — a standard Newton solver can be trapped in either local minimum. Right: with our quality regularization, the suboptimal local minimum is suppressed while the global minimum is preserved. Both panels share the same color scale; insets show 3D side views of the landscape. The unchanged minimum at configuration 2 confirms that our regularizer selectively penalizes badly-shaped configurations without biasing physically meaningful solutions.}
    \label{fig:ip_landscape}
\end{figure}

\section{Resolving Variational ITR Pathologies}
\label{sec:quality}
When exploring ITR for thin shells, we identify two distinct pathologies that fundamentally challenge variational r-adaptivity: the non-convexity of the spatially discretized IP, which admits multiple non-equivalent local minima that trap the solver in suboptimal configurations (Section~\ref{sec:basin_escape}); and the spurious low-energy states produced by the variational remesher exploiting degenerate elements, where the FE approximation breaks down (Section~\ref{sec:degeneracy}). These two pathologies are distinct in origin but often co-occur in practice. We construct a single quality regularization term that simultaneously resolves both (Section~\ref{sec:quality_energy}).

\subsection{Suboptimal Local Minima}
\label{sec:basin_escape}
The spatially discretized IP is inherently non-convex, an unavoidable consequence of discretization combined with contact constraints and large-deformation elasticity. While the spatially continuous IP can also admit multiple local minima, corresponding to physically valid equilibria such as different stable wrinkle patterns in cloth draping, discretization introduces \emph{additional} non-equivalent local minima that are not physically meaningful. These are suboptimal (lower-quality) basins in the optimization sense (higher discrete IP under the same boundary conditions) and represent artifacts of the discretization rather than distinct physical solutions. Once a Newton solver becomes trapped in such a suboptimal basin, escaping to a superior basin is computationally difficult: descent-based solvers (including Newton-type methods employed) lack  practical mechanisms to cross over local energy barriers.

We illustrate this phenomenon using a simple 2D rod simulation (Figure~\ref{fig:ip_landscape}). The discrete energy landscape exhibits two distinct local minima differing in physical configuration: the global minimum corresponds to the rod bending compliantly over the contact corner, while the suboptimal minimum leaves one DOF redundant on the flat region, producing a stiffer discretization. Both basins are attractive to a standard Newton solver, and which one is reached depends entirely on initialization.
Such configurations are pervasive in practical simulations. Once trapped, the solver tends to create entangled mesh as simulation proceeds (Figure~\ref{fig:bunching}), compounding the issue. We empirically observe that in these problematic configurations, the suboptimal minima are typically supported by, and strongly correlated with, degraded element quality (e.g., configuration 1 in Figure~\ref{fig:ip_landscape} involves an over-stretched segment relative to the global minimum's discretization). This empirical correlation suggests that a quality-based regularizer can selectively penalize the basins associated with bad mesh configurations, enabling escape from these suboptimal minima. Simple mesh-level interventions such as local Laplacian smoothing, small-element barriers~\cite{Wen:2025:ORI}, or other isotropy-seeking metrics, cannot reshape the whole IP landscape to facilitate escape. A more sophisticated regularization is needed: one that actively biases the optimization away from suboptimal basins while preserving the global minimum.

\subsection{Degenerate Elements}
\label{sec:degeneracy}
The second, distinct pathology arises particularly from the use of low-order finite elements in the variational ITR formulation. It is well established that for simplicial elements with piecewise-linear approximations, accuracy is fundamentally governed by element size and shape~\cite{Shewchuk:2002:WIA}, and element quality dictates the conditioning of stiffness matrices and the convergence rate of FEM~\cite{Ferguson:2023:ITR}. Compared to fixed-mesh simulation, r-adaptive solvers redistribute vertices to minimize the total objective, typically by clustering DOFs in regions of high deformation or generating anisotropic elements. Such redistribution is desirable when it efficiently captures strain distribution. However, when element quality deteriorates beyond a certain point, the discrete FE approximation locally loses fidelity~\cite{Shewchuk:2002:WIA}, and the IP value computed on degenerate elements ceases to be a reliable proxy for the true continuous energy. The optimization can then exploit this discretization error: the solver finds artificially low IP values not by improving the physical solution, but by collapsing elements into degenerate configurations.

\begin{figure}[!t]
    \centering
    \includegraphics[width=\linewidth]{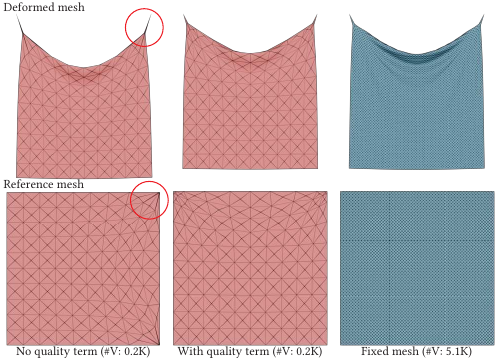}
    \vspace{-8mm}
    \caption{\textbf{Necessity of the quality term.} Without the quality regularization~\cite{Wen:2025:ORI}, the solver exploits mesh distortion (highlighted in circles) to reach an unphysically low energy of $-1.21\times 10^{-4}$. By enabling the quality term (our method), the r-adaptive solution avoids this degeneracy and converges to $-1.13 \times 10^{-4}$, a value consistent with the high-resolution fixed mesh benchmark ($25.5\times$ more vertices).}
    \vspace{-6mm}
    \label{fig:cheat}
\end{figure}

We analyze a didactic draping scenario to illustrate this exploitation (Figure~\ref{fig:cheat}). The discrete elastic potential is $\Sigma_{s\in\mathcal{T}}a_s(u)\Psi_s(x,u)$, where $\Psi_s$ is the per-element energy density and $a_s(u)$ is the rest area. Elastic restoring forces scale with $a_s(u)$, while the gravitational load remains constant under mass conservation. Consequently, as the solver shrinks $a_s(u)$ toward zero, the affected elements effectively "soften": to balance gravity, they must undergo larger deformations, causing the cloth to sag excessively and lowering the gravitational potential energy. This creates a spurious incentive for the solver to collapse elements toward zero area: artificially lower IP values are achieved without any physical improvement. In the unregularized limit, the cloth produces artificial sharp seams and sags below even a high-resolution fixed-mesh benchmark with $25.5\times$ more vertices.
This pathology is fundamentally distinct from the suboptimal local minima problem in Section~\ref{sec:basin_escape}, both in mechanism and in failure direction. In Section~\ref{sec:basin_escape}, the spatially discrete IP itself is non-convex and the solver becomes trapped at \emph{higher} IP values (suboptimal local minima above the global minimum), but the IP remains a reliable proxy for physical quality. Here, the FE approximation underlying the IP becomes locally invalid, and the solver achieves \emph{artificially lower} IP values through cheating modes that bear no relation to physical improvement. The two pathologies often co-occur (\textit{element bunching} in Figure~\ref{fig:bunching}): a solver trapped in a suboptimal basin tends to drive elements toward degeneracy, and degenerate elements support new spurious minima, but they require distinct theoretical understanding.

\subsection{Element Quality Energy}
\label{sec:quality_energy}
We now construct a single regularization term that simultaneously resolves both pathologies. The two requirements are: (a) reshape the discrete IP landscape to suppress discretization-induced suboptimal basins (Section~\ref{sec:basin_escape}), and (b) prevent the solver from exploiting degenerate elements through cheating modes (Section~\ref{sec:degeneracy}). A naive approach is to add an isotropy-seeking quality term that uniformly penalizes anisotropic elements. However, r-adaptive schemes inherently rely on healthy anisotropy to capture deformation details~\cite{Wen:2025:ORI}; isotropy-seeking penalties counteract the r-adaptive objective and bias the solver away from physically optimal solutions. Our term must therefore act as a \emph{selective barrier}: it must permit healthy anisotropy while activating only as elements approach an entangled or degenerate state.

The specific quality measure can be tailored to the geometric primitive. For the rod example in Figure~\ref{fig:ip_landscape}, we use a segment-length-based quality measure that activates as segments deviate substantially from their initial lengths. For thin shells, where elements are 2D triangles and both anisotropy and area distortion are relevant, we propose a measure based on reference triangle distortion.  For each triangle, let $F$ be the deformation gradient mapping the initial reference configuration to the current configuration, with singular values $\sigma_1, \sigma_2$. We use the MIPS~\cite{Hormann:2000:MIPS} energy density $\Psi_{\text{MIPS}} = \frac{\sigma_1}{\sigma_2} + \frac{\sigma_2}{\sigma_1}$,
which satisfies $\Psi_{\text{MIPS}}\geq 2$ with equality when $F$ is conformal. Departures from conformality (anisotropy or shear) increase $\Psi_{\text{MIPS}}$ continuously. We define the quality energy with an activation threshold $\epsilon_q$ (typically $\epsilon_q=2.5$):
\begin{equation}\label{eq:quality}
E_{\text{quality}} = \max((\Psi_{\text{MIPS}} - \epsilon_q)^3, 0).
\end{equation}

This element quality energy $E_{\text{quality}}$ enters the remeshing objective as the regularization term $C(u)$ introduced in Section~\ref{sec:itr_framework}, summed over all triangles in the mesh. Throughout this work, we set its weight equal to that of the inertial energy (i.e., unit weight), maintaining dimensional consistency with the inertial scaling without per-scene tuning; we therefore omit the explicit coefficient in Eq.~(\ref{eq:quality}). 
The cubic penalization is chosen for $C^2$-continuity at the activation threshold. Below $\epsilon_q$, the regularizer is inactive and the variational remesher freely generates healthy local anisotropies that capture physical detail; above $\epsilon_q$, the penalty grows cubically, sharply penalizing highly distorted triangles. Critically, adding this term to the objective reshapes the discrete IP landscape: it lifts the energy of basins supported by degraded mesh quality, enabling the solver to escape suboptimal physical minima even when those minima are not themselves degenerate (Figure~\ref{fig:ip_landscape}, right). While $\Psi_{\text{MIPS}}$ is conformal and does not directly penalize small elements, the fixed boundary of the reference mesh provides an implicit safeguard against degeneracy: collapsing any interior element forces its neighbors to undergo large distortion to preserve total area, sharply increasing their $\Psi_{\text{MIPS}}$ values and triggering the regularizer. Although MIPS itself is scale-invariant, this boundary-preservation mechanism gives our quality term effective control over both element shape and area.

\begin{figure*}[!t]
    \centering
    \includegraphics[width=1.0\linewidth]{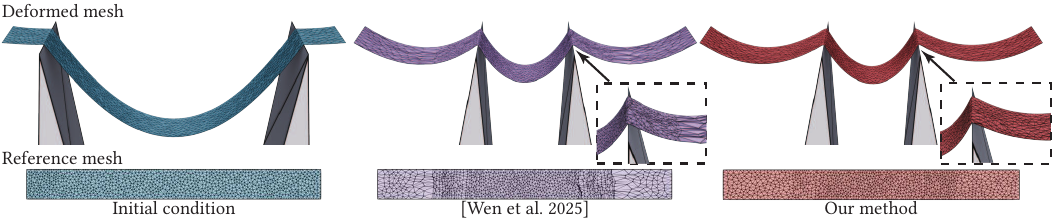}
    \caption{\textbf{Efficacy of the quality regularization term.} We simulate a discretized strip, pinned at both ends, interacting with rigid obstacles. For each panel, the upper sub-row displays the deformed mesh (world space) and the lower sub-row shows the reference mesh (material space). \textbf{Left:} the initial configuration. \textbf{Middle:} without quality regularization~\cite{Wen:2025:ORI}, the r-adaptive solver exhibits the \emph{element bunching} failure mode: it minimizes energy via uncontrolled vertex clustering along the sharp contact edges, resulting in severe mesh tangling and degenerate elements (see inset). \textbf{Right:} with our quality term, the solver maintains a well-conditioned triangulation while still adapting to the geometric deformation. Note that our method preserves the sharp contact interfaces while ensuring numerical stability.}
    \label{fig:bunching}
\end{figure*}

By employing $\Psi_{\text{MIPS}}$, we implicitly assume the initial mesh has high quality; all examples in this paper are initialized with Delaunay triangulation or structured grid. A simpler alternative would be to revert the reference mesh to its initial state whenever quality degrades, but this introduces large $L^2$ projection errors and temporal instability. Our framework is metric-agnostic: $\Psi_{\text{MIPS}}$ can be substituted with ODT energy, minimum angle constraints, aspect ratios, or other distortion measures. We illustrate the practical effect of our regularizer in Figure~\ref{fig:bunching}, where the original r-ITR~\cite{Wen:2025:ORI} solver becomes trapped in a degenerate, low-quality configuration during sliding contact, while our quality energy enables stable, well-conditioned remeshing throughout and faithfully captures the wrinkle transitions near the contact edge (also observed in high-resolution fixed-mesh simulations in the video).

\section{Dynamic ITR Solver}
\label{sec:acceleration}
Computational efficiency remains the primary barrier for adopting In-Timestep Remeshing (ITR) in both h-~\cite{Ferguson:2023:ITR} and r-adaptivity~\cite{Wen:2025:ORI}. The full r-adaptive ITR framework (Equation~\ref{eq:optimization}) is a bi-level optimization with two distinct performance bottlenecks. First, the inner nonlinear timestep optimization is repeatedly invoked during outer line search, and tightly solving it at every outer iteration is unnecessarily expensive. Second, evaluating gradients and Hessians with respect to reference coordinates $u$ — particularly the $L^2$ projection of inertial quantities, which requires dense per-element quadrature~\cite{Wen:2025:ORI} — dominates per-iteration cost in elastodynamics. We address both bottlenecks with strategies that exploit the structure of the bi-level optimization itself: 1) inner solve accuracy can be progressively tightened as the outer loop converges (Section~\ref{sec:tolerance}); and 2) derivatives can be carefully reused across iterations, exploiting the within-timestep coherence of the ITR r-adaptive mesh updates (Section~\ref{sec:coherence}). Each strategy independently provides measurable speedup, and combining them yields significant acceleration; to understand their impact we also provide an ablation isolating their respective contributions in Figure~\ref{fig:performance}. Although our primary focus is variational r-adaptive cloth simulation, both strategies apply equally to volumetric ITR, providing substantial gains in both domains.

\subsection{Dynamic Tolerance Adjustment}
\label{sec:tolerance}
\setlength{\columnsep}{2mm}
\begin{wrapfigure}{r}{0.6\linewidth}
    \centering
    \vspace{-3mm}    
    \includegraphics[width=0.8\linewidth]{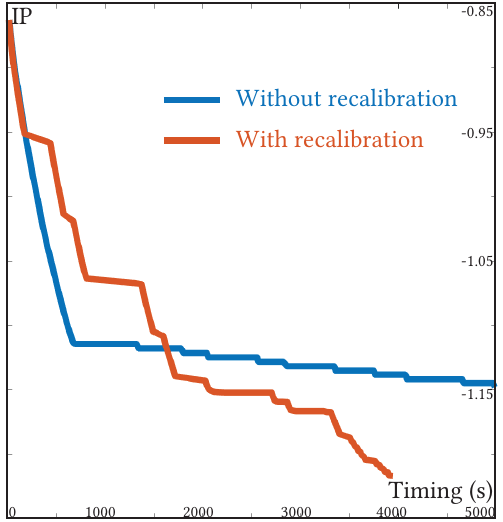}
    \vspace{-4mm}    
    \caption{\textbf{Periodic Recalibration Impact.} Without recalibration (blue), accumulated drift from loose inner solves stalls outer convergence. Our recalibration (orange) periodically enforces strict-tolerance solve, restoring valid descent and continued energy reduction.}
    \vspace{-5mm}    
    \label{fig:calibration}
\end{wrapfigure}
Maintaining a strict inner tolerance, and so high-accuracy objective evaluations, is computationally redundant when the outer solve is far from convergence~\cite{Pedregosa:2016:HOW, Bhatia:2025:PRDP}. We progressively tighten our inner tolerance with exponential decay scheduling. Letting $\epsilon_x^0$ denote an initially relaxed inner-solve tolerance and $\epsilon_x^*$ our final (target) strict tolerance, we dynamically refine the inner tolerance at outer iteration $i$ by $\epsilon_x^i = \max(\lambda \epsilon_x^{i-1}, \epsilon_x^*)$,
where $\lambda \in (0,1)$ controls the decay rate. We use $\lambda=0.98, \epsilon_x^0=0.01$, and $\epsilon_x^*=10^{-4}$ across all examples. While this scheduling strategy already gains significant speedups for many timestep solves, on its own we observe it is insufficient in challenging simulations with high-speed dynamics, complex contact, and/or large deformations. In these cases accumulated inaccuracies across repeated loose inner solves can eventually violate outer-solve descent conditions (Figure~\ref{fig:calibration}).
To address this, we periodically (re-)apply the strict final tolerance $\epsilon_x^*$ at a fixed frequency (every $k=10$ iterations). This realigns the solver with the accurate energy landscape, providing progress even in challenging simulation settings. See pseudocode in our supplemental for details.

\subsection{Exploiting Within-Timestep Coherence}
\label{sec:coherence}
Because r-adaptivity preserves mesh connectivity, ITR optimization produces incrementally changing, locally clustered vertex updates across outer iterations. Re-computing all stencils' (expensive) derivatives from scratch each iteration is then redundant. We exploit this within-timestep coherence to selectively cache and reuse derivatives. At outer iteration $j$, we cache the computed element-wise derivatives alongside the corresponding vertex configuration $(x_j,u_j)$. At iteration $j+1$, before computing the Newton direction, we measure per-vertex displacement $d_i = \max(\|u_i^{j+1} - u_i^j\|, \|x_i^{j+1} - x_i^j\|)$ as a measure of need for update.
For each stencil, if maximum $d_i$ over its vertices is below threshold, $\epsilon_d$, 
we treat it as coherent and retrieve cached derivatives; otherwise, we mark the stencil active and recompute its derivatives. To guarantee progress across flat energy landscape regions, we also re-evaluate the top 10\% of stencils with the largest relative displacements at every iteration, regardless of activity status. We define the \emph{active rate} as the fraction of active stencils per iteration. As shown in Figure~\ref{fig:performance}, the active rate remains consistently low across both volumetric and cloth simulations, yielding up to $5\times$ speedup over full re-evaluation.

\section{Evaluation}
We implement our method in C++ using Intel TBB for parallelism, Eigen~\cite{eigenweb} for linear algebra and sparse Cholesky ($LL^T$) factorization, and the Triangle library~\cite{Shewchuk:1996:TRI} for Delaunay triangulation. All experiments run on a MacBook Pro with Apple M3 Pro and 18 GB unified memory.

\paragraph{Parameter settings} All algorithmic parameters are fixed across the evaluation with $\epsilon_x^* = 10^{-4}$, $\lambda = 0.98$, $\epsilon_q = 2.5$, $\epsilon_d = 10^{-4}$, and $k = 10$. We choose $\epsilon_q=2.5$ to leave a reasonable tolerance band above the conformal lower bound $\Psi_{\text{MIPS}}=2$, keeping the regularizer inactive for nearly-conformal elements and activating only as non-conformal distortion grows substantially; $\lambda=0.98$ balances progressive inner-tolerance tightening with sufficient relaxation between recalibrations. The weight of the quality regularization is set equal to the inertial energy weight for dimensional consistency. Sensitivity analysis for $\epsilon_q$, and material parameters for all examples are provided in the supplemental.


\paragraph{Evaluation protocol.} Thin-shell simulations under contact are highly nonconvex, and their solutions generally do not converge to a unique configuration under mesh refinement — different resolutions and discretizations can produce wrinkles and folds with different frequencies and shapes, all of which are physically valid local minima of the incremental potential. We therefore do not evaluate against a fixed ``ground truth'' configuration. Instead, we assess our method along three axes: (1) \emph{energy reduction} at equal vertex budget, demonstrating that our $r$-adaptive solver finds lower-energy configurations than fixed meshes of the same resolution; (2) \emph{conforming contact adaptation}, where vertices automatically cluster toward sharp contact features to produce crisp, well-resolved contact interfaces; and (3) \emph{reduction of locking artifacts}, comparing against fixed-mesh baselines that exhibit characteristic stiffness-induced bulging and over-coarse fold patterns.

\subsection{Energy Reduction at Equal Budget}
In the dynamic mannequin example (Figure~\ref{fig:teaser}, left), the per-frame IP reduction plot shows our method achieves the largest reductions during the most intense deformation phases (frames 40--100). For static contact scenarios, in the two-point draping experiment (Figure~\ref{fig:two-point-draping}), our method reaches lower potential energy than the fixed mesh baseline with significantly fewer vertices and reduced computation time. This advantage remains consistent across varying material parameters: in the cube draping test (Figure~\ref{fig:cube_draping}), our method maintains a steady energy descent across different cloth thicknesses, reaching a lower converged energy state than the fixed mesh at equivalent vertex budget.

\subsection{Conforming Contact Adaptation}
\begin{figure}[!t]
    \centering
    \includegraphics[width=1.0\linewidth]{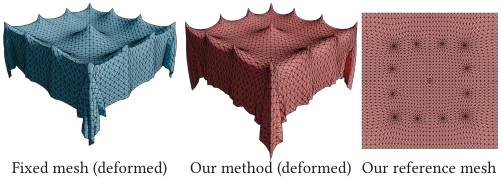}
    \vspace{-6mm}
    \caption{\textbf{Draping simulation on rigid spikes.} We compare a fixed mesh simulation (left) against our r-adaptive method (middle); right panel shows our method's reference mesh in material space. Both simulations start from a symmetric, structured initial reference mesh. Our method automatically concentrates resolution at high-curvature contact regions (spike tips), preserving smooth wrinkles, sharp contact geometry, and problem symmetry throughout. The fixed mesh, restricted to its uniform discretization, suffers from undersampling and aliasing artifacts.}
    \label{fig:spikes}
\end{figure}

\begin{figure*}
    \centering
    \includegraphics[width=1.0\linewidth]{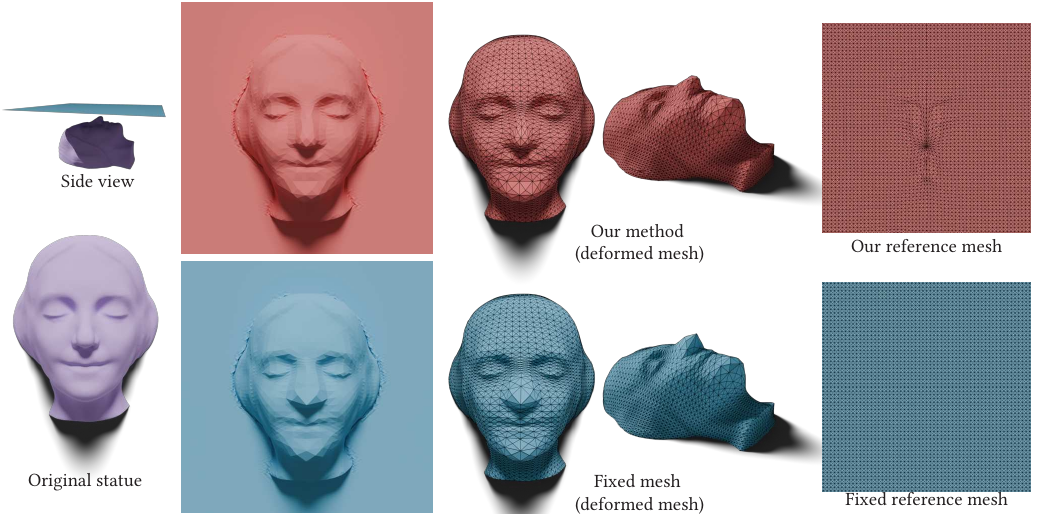}
    \caption{\textbf{Extreme draping over "L'Inconnue de la Seine".} An ultra-thin film ($h = 0.01$ mm) drapes under gravity over a complex statue. Left: original statue and initial setup. Top: our r-adaptive method captures intricate facial features (eyebrows, lips, jawline); the reference mesh (top right) reveals a striking "ghost face" as the solver automatically concentrates resolution toward high-curvature contact regions (eyes, nose, mouth) — emerging purely from variational optimization without explicit knowledge of underlying geometry. Bottom: the fixed mesh suffers severe undersampling, producing a distorted nose and washed-out details at equivalent vertex budget.}
    \label{fig:seine}
\end{figure*}

Unlike stiff volumetric objects, cloth deforms compliantly against external obstacles, making conforming contact adaptation particularly important. In Figure~\ref{fig:spikes}, the fixed mesh's uniform discretization fails to capture the sharp geometry of the underlying spikes. Our method clusters vertices toward high-curvature regions, capturing sharp features and forming smooth, symmetric wrinkle contours. In Figure~\ref{fig:cube_draping}, vertices automatically align with the sharp contact line of the box, allowing the cloth to drape lower and reach a lower energy state. In the L'Inconnue de la Seine draping (Figure~\ref{fig:seine}), the fixed mesh's uniform sampling produces a distorted nose and blurred jawline, while our method captures fine geometric details and reveals a ``ghost face'' in the reference mesh through emergent vertex clustering. 

\subsection{Reduction of Locking Artifacts}
Membrane locking — artificial stiffening of bending modes that arises from low-order membrane elements, is a well-known limitation of 
shell models
that our r-adaptive method cannot avoid (as it employs 
the same underlying shell discretization as fixed-mesh methods).
However, our r-adaptive method substantially reduces locking-induced artifacts by redistributing degrees of freedom to optimal positions, allocating resolution where deformation modes are most prone to over-stiffening.

\begin{figure}[!b]
    \centering
    \includegraphics[width=1.0\linewidth]{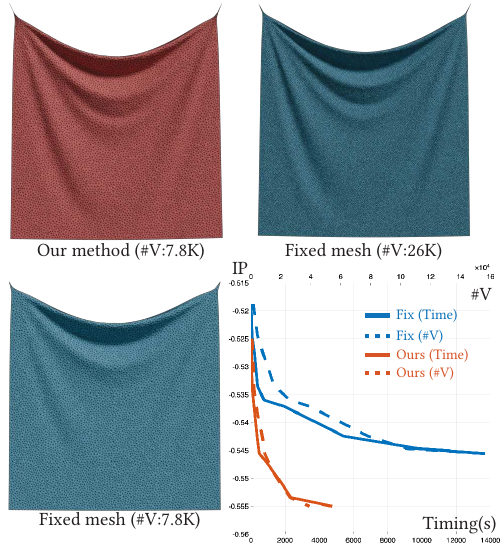}
    \vspace{-8mm}    
    \caption{\textbf{Two-point fixed cloth draping.} The visual comparison demonstrates that our method (Top Left) faithfully reproduces the rich buckling behavior of a high-resolution reference (Top Right, 26K vertices) using only 7.8K vertices. In contrast, a standard fixed mesh at the same resolution (Bottom Left) appears overly stiff and lacks detail. (Bottom Right) Quantitative analysis confirms this advantage: the energy plot shows that our approach (orange curves) outperforms the fixed mesh baseline (blue curves) in both convergence speed and spatial discretization, allowing us to reach lower converged energy states with significantly fewer vertices and shorter time.}
    \label{fig:two-point-draping}
\end{figure}

\begin{figure}[!b]
    \centering
    \includegraphics[width=1.0\linewidth]{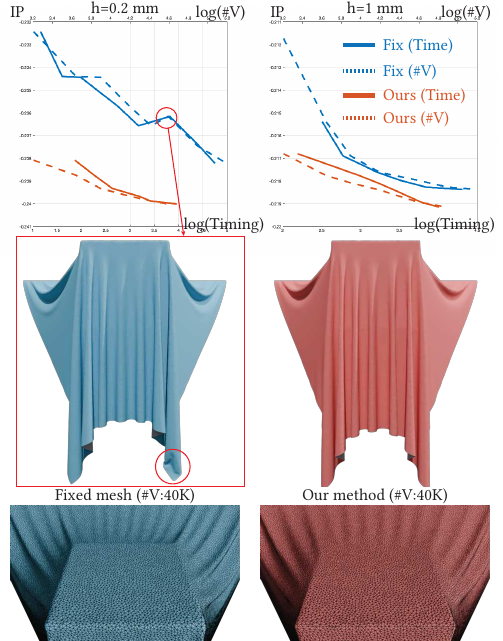}
    \vspace{-7mm}    
    \caption{\textbf{Efficiency and fidelity in cube draping.} Top: performance plots for two cloth thicknesses ($h=0.2$ mm and $h=1$ mm) showing our r-adaptive method (red) consistently reaches lower potential energy with fewer vertices and less computation time than the fixed mesh (blue). Middle: visual comparison for the thin cloth ($h=0.2$ mm). The fixed mesh (left) suffers from membrane locking — corresponding to the non-monotonic energy jump (red circle) — producing asymmetric wrinkles and non-conforming contacts at the corners. Our method (right) resolves these artifacts. Bottom: zoomed wireframes show vertices automatically aligning with the sharp contact line of the box, forming clean, symmetric high-frequency wrinkles.}
    \label{fig:cube_draping}
\end{figure}

In the dynamic example shown on the left of Figure~\ref{fig:teaser}, we simulate a dress driven by a mannequin performing extremely large, high-speed motions. As the mannequin executes a rapid kick, the fixed mesh exhibits pronounced locking artifacts, produces aliased wrinkle patterns and fails to capture intricate garment dynamics. Our method resolves smooth wrinkle flows, preserves complex multi-layered folds, and produces natural draping as the motion settles without long locked creases along the bottom. In the static example (Figure~\ref{fig:teaser}, right), a square cloth drapes over a sphere: the fixed mesh produces blocky, stiffness-locked folds, while our method generates smooth, continuous folds that conform to the underlying geometry. The mean curvature distribution provides quantitative support: the fixed mesh exhibits more high-curvature concentration in the right tail of $\log(H^2)$ — a signature of locking-induced sharp folds — while our method shifts the distribution toward smoother, lower curvature folds. Notably, neither our regularizer nor the underlying volumetric-reduced shell model directly penalizes mean curvature, supporting the interpretation that this reduction reflects locking alleviation rather than objective design.
In the two-point fixed cloth draping (Figure~\ref{fig:two-point-draping}), a 7.8K fixed mesh appears overly stiff and lacks wrinkle details, while our method reproduces the rich buckling behavior of a 26K mesh using $3.3\times$ fewer vertices. In the cube draping test (Figure~\ref{fig:cube_draping}), the fixed mesh exhibits non-monotonic energy jumps indicative of locking, which our method avoids entirely. 

\subsection{Solver Performance and Speedup}
\begin{figure}[!t]
    \centering
    \includegraphics[width=1.0\linewidth]{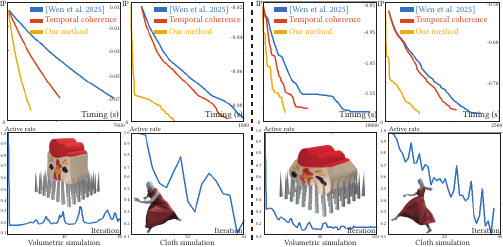}
    \vspace{-6mm}
    \caption{\textbf{Convergence and efficiency analysis.} Solver performance on representative frames (insets) under mild (left) and large (right) deformations, on volumetric (\#V: 53K) and cloth (\#V: 14K) simulations. \textbf{Top}: per-frame IP reduction vs wall-clock time, comparing baseline r-ITR~\cite{Wen:2025:ORI} (blue), within-timestep coherence only (orange), and our full method (yellow). Each strategy independently improves convergence; combined, they yield ~$5\times$ and ~$3\times$ speedup over baseline for volumetric and cloth respectively. \textbf{Bottom}: active rate (fraction of stencils requiring re-computation) over outer iterations. The sharp early decay confirms localized derivative updates with consistently low active rates regardless of deformation intensity.}
    \label{fig:performance}
\end{figure}

The ablation in Figure~\ref{fig:performance} confirms that each acceleration strategy contributes independently, with the combination yielding around $5\times$ and $3\times$ speedup over the baseline r-ITR~\cite{Wen:2025:ORI} for volumetric and cloth simulations respectively. We further verified our solver on a 3D volumetric example (the masticator experiment from~\cite{Wen:2025:ORI}; see supplemental), achieving $6\times$ speedup over the full animation sequence with average Hausdorff distance to a fine-mesh benchmark increasing by only $1\%$.

\section{Conclusion}
We present the first r-adaptive method for frictionally contacting cloth simulation compatible with modern IPC-based pipelines, identifying and addressing the fundamental obstacles that prevented previous ITR-based methods from working for thin shells. Our key technical contributions are a degeneracy-activated quality regularization that resolves both the IP nonconvexity and FE-degeneracy pathologies of variational r-adaptivity in thin shells, and a temporally adaptive solver that accelerates in-timestep remeshing by $3\text{-}8\times$ (for both shell and volumetric FE simulations). Our method also highlights r-adaptivity's unique advantages in contact-rich scenarios: by moving vertices in the undeformed domain, contact features are well-captured (e.g., over sharp bends) in the deformed mesh. At the same time, we address a fundamental cloth-specific r-adaptive failure mode: \textit{element bunching} during sliding contact over sharp features. Together, these contributions enable cloth simulations to resolve folds, wrinkles, and sharp contact regions with significantly higher fidelity than fixed-resolution meshes under equal budgets.

Our current formulation has several limitations: it assumes flat undeformed configurations (extension to curved shells is open), does not introduce additional DOFs for boundary contacts (limiting edge contact resolution), and is demonstrated on specific FEM cloth solvers~\cite{Chen:2018:PSO,Wen:2023:KLS}. Promising directions include extension to one-dimensional rods and intrinsically curved shells, and combining r-adaptivity with selective h-adaptivity for next-generation hybrid adaptive simulation.

\bibliographystyle{ACM-Reference-Format}
\bibliography{references}

\end{document}


\setcopyright{acmlicensed}
\acmJournal{TOG}
\acmYear{2025} \acmVolume{44} \acmNumber{4} \acmArticle{} \acmMonth{8}\acmDOI{10.1145/3731204}

\title{Supplement to Variational r-Adaptive Cloth Simulation}

%
%




\maketitle

\section{Cloth Elasticity Derivatives}
\label{app:derivatives}
Following~\cite{Wen:2023:KLS} and assuming the rest shape is completely flat (mean curvature $H=0$ and Gauss curvature $K=0$ everywhere), the elasticity per triangle $\triangle ijk$:
\begin{equation}
E_{ijk}=\frac{1}{2}hA_{ijk}\psi
\end{equation}
where
\begin{equation}
\begin{aligned}
\psi&= \psi(C_1)+\psi(C_2)\\
C_1&= F_0+\frac{\sqrt{3}}{6}hF_1,\quad C_2=F_0-\frac{\sqrt{3}}{6}hF_1\\
F_0&=tT^{-1}, \quad F_1=qT^{-1}\\
t&=\begin{bmatrix}x_j-x_i & x_k-x_i & n_{ijk} \end{bmatrix}\\
q&=2\begin{bmatrix}n_i-n_j & n_i-n_k & 0 \end{bmatrix}\\
T&=\begin{bmatrix}u_j-u_i & u_k-u_i & 0\\ 0 & 0& 1 \end{bmatrix}\\
\end{aligned}
\end{equation}

Here, $n$ is the mid-edge normal. Regarding the gradient,
\begin{equation}
\begin{aligned}
\frac{\partial E}{\partial x}&=\frac{1}{2}hA\frac{\partial \psi}{\partial x}\\
\frac{\partial E}{\partial u}&=\frac{1}{2}h(\frac{\partial A}{\partial u}\psi + A\frac{\partial \psi}{\partial u})
\end{aligned}
\end{equation}

where 
\begin{equation}
\begin{aligned}
\frac{\partial \psi}{\partial x} &= (\frac{\partial \psi_1}{\partial C_1}+\frac{\partial \psi_2}{\partial C_2})\frac{\partial F_0}{\partial t}\frac{\partial t}{\partial x}+\frac{\sqrt 3 h}{6}(\frac{\partial \psi_1}{\partial C_1}-\frac{\partial \psi_2}{\partial C_2})\frac{\partial F_1}{\partial q}\frac{\partial q}{\partial x}\\
\frac{\partial \psi}{\partial u}&=(\frac{\partial \psi_1}{\partial C_1}+\frac{\partial \psi_2}{\partial C_2})\frac{\partial F_0}{\partial T}\frac{\partial T}{\partial u}+\frac{\sqrt3h}{6}(\frac{\partial \psi_1}{\partial C_1}-\frac{\partial \psi_2}{\partial C_2})\frac{\partial F_1}{\partial T}\frac{\partial T}{\partial u}
\end{aligned}
\end{equation}

Regarding the Hessian,
\begin{equation}
\begin{aligned}
\frac{\partial^2 E}{\partial x^2} &= \frac{1}{2}hA\frac{\partial^2 \psi}{\partial x^2} \\
\frac{\partial^2 E}{\partial u^2} &=\frac{1}{2}h(\frac{\partial^2 A}{\partial u^2}\psi + \frac{\partial A}{\partial u}\frac{\partial \psi}{\partial u}^T + \frac{\partial \psi}{\partial u}\frac{\partial A}{\partial u}^T + A\frac{\partial^2 \psi}{\partial u^2} ) \\
\frac{\partial^2 E}{\partial x \partial u} &=\frac{1}{2}h (\frac{\partial \psi}{\partial x} \frac{\partial A}{\partial u}^T + A\frac{\partial^2 \psi}{\partial x \partial u} )
\end{aligned}
\end{equation}

where
\begin{equation}
\begin{aligned}
\frac{\partial^2 \psi}{\partial x^2} &= (\frac{\partial^2 \psi_1}{\partial C_1^2}\frac{\partial C_1}{\partial x}+\frac{\partial^2 \psi_2}{\partial C_2^2}\frac{\partial C_2}{\partial x})\frac{\partial F_0}{\partial t}\frac{\partial t}{\partial x}\\&+(\frac{\partial \psi_1}{\partial C_1}+\frac{\partial \psi_2}{\partial C_2})\frac{\partial F_0}{\partial t}\frac{\partial^2 t}{\partial x^2}\\
&+\frac{\sqrt3h}{6}(\frac{\partial^2 \psi_1}{\partial C_1^2}\frac{\partial C_1}{\partial x}-\frac{\partial^2 \psi_2}{\partial C_2^2}\frac{\partial C_2}{\partial x})\frac{\partial F_1}{\partial q}\frac{\partial q}{\partial x} \\&+\frac{\sqrt3h}{6} (\frac{\partial \psi_1}{\partial C_1}-\frac{\partial \psi_2}{\partial C_2}) \frac{\partial F_1}{\partial q} \frac{\partial^2 q}{\partial x^2}
\end{aligned}
\end{equation}

\begin{equation}
\begin{aligned}
\frac{\partial^2 \psi}{\partial u^2}&=(\frac{\partial^2 \psi_1}{\partial C_1^2}\frac{\partial C_1}{\partial u}+\frac{\partial^2 \psi_2}{\partial C_2^2}\frac{\partial C_2}{\partial u})\frac{\partial F_0}{\partial u}\\&+(\frac{\partial \psi_1}{\partial C_1}+\frac{\partial \psi_2}{\partial C_2})\frac{F_0}{\partial T^{-1}}\frac{\partial T}{\partial u}^T\frac{\partial^2 T^{-1}}{\partial T^2}\frac{\partial T}{\partial u}\\
&+\frac{\sqrt3h}{6}(\frac{\partial^2 \psi_1}{\partial C_1^2}\frac{\partial C_1}{\partial u}-\frac{\partial^2 \psi_2}{\partial C_2^2}\frac{\partial C_2}{\partial u})\frac{\partial F_1}{\partial u}\\&+\frac{\sqrt3h}{6}(\frac{\partial \psi_1}{\partial C_1}-\frac{\partial \psi_2}{\partial C_2})\frac{\partial F_1}{\partial T^{-1}}\frac{\partial T}{\partial u}^T\frac{\partial^2 T^{-1}}{\partial T^2}\frac{\partial T}{\partial u}
\end{aligned}
\end{equation}

\begin{equation}
\begin{aligned}
\frac{\partial^2 \psi}{\partial x\partial u}&=(\frac{\partial^2 \psi_1}{\partial C_1^2}\frac{\partial C_1}{\partial u} + \frac{\partial^2 \psi_2}{\partial C_2^2}\frac{\partial C_2}{\partial u}) \frac{\partial F_0}{\partial x}\\&+(\frac{\partial \psi_1}{\partial C_1}+\frac{\partial \psi_2}{\partial C_2})\frac{\partial t}{\partial x}^T\frac{\partial^2 F_0}{\partial t\partial T^{-1}}\frac{\partial T^{-1}}{\partial u} \\
&+\frac{\sqrt3h}{6}(\frac{\partial^2 \psi_1}{\partial C_1^2}\frac{\partial C_1}{\partial u} - \frac{\partial^2 \psi_2}{\partial C_2^2}\frac{\partial C_2}{\partial u})\frac{\partial F_1}{\partial x} \\&+\frac{\sqrt3h}{6}(\frac{\partial \psi_1}{\partial C_1}-\frac{\partial \psi_2}{\partial C_2}) \frac{\partial q}{\partial x}^T \frac{\partial^2 F_1}{\partial q \partial T^{-1}}\frac{\partial T^{-1}}{\partial u} 
\end{aligned}
\end{equation}

\section{Supplemental Algorithms, Tables, and Figures}
\begin{algorithm}
\caption{Dynamic Tolerance Adjustment}\label{alg:exp_decay}
  \begin{algorithmic}[1]
    \Procedure{R-ITR}{$x^t,v^t, u^t, e$}
    \State $\tilde{x}^t \gets x^t + \Delta tv^t$
    \State $x \gets \argmin_{x} E_t\big(x, u^t\big)$ \textcolor{ForestGreen}{subject to $\epsilon_x^*$}
    \State $u \gets u^t$
    \State \textcolor{ForestGreen}{$\epsilon_x \gets \epsilon_x^0, \,\text{iter}\gets 0$}
    \Loop
    \State $\Delta u \gets -H^{-1}g$
    \State  \textbf{if} $||\Delta u|| \leq \epsilon_u$ \textbf{break} 
    \State $\alpha \gets \min\big(1, \text{StepFilter}(u,\Delta u)\big)$
    \State $R_0 \gets R(x, u)$
    \State \textcolor{ForestGreen}{$\epsilon_x \gets \max(\lambda\epsilon_x, \epsilon_x^*)$}
    \Do
    \State $u' \gets u + \alpha \Delta u$
    \State $x \gets \argmin_{x} E_t\big(x,u'\big)$ \textcolor{ForestGreen}{subject to $\epsilon_x$}
    \State $\alpha \gets \alpha/2$
    \doWhile $R(x,u') > R_0$
    \State $u \gets u'$
    \textcolor{ForestGreen}{\If{$\text{iter} \equiv 0 \pmod k$}
    \State $x \gets \argmin_{x} E_t\big(x, u  \big)$ \textcolor{ForestGreen}{subject to $\epsilon_x^*$}
    \State $\epsilon_x\gets \epsilon_x^0$
    \EndIf
    \State $\text{iter} \gets \text{iter+1}$}
    \EndLoop
    \State $x \gets \argmin_{x} E_t\big(x, u^t\big)$ \textcolor{ForestGreen}{subject to $\epsilon_x^*$}    
    \State $u^{t+1} \gets  u, x^{t+1} \gets x$
    \State $v^{t+1} \gets 1/\Delta t \big(x^{t+1} - \pi_t\big(x^t,(u^{t+1}, e)\big) \big)$
    \State \Return $x^{t+1}, v^{t+1}, u^{t+1}$
    \EndProcedure
  \end{algorithmic}
\end{algorithm}

\begin{figure}[!h]
    \centering
    \includegraphics[width=1.0\linewidth]{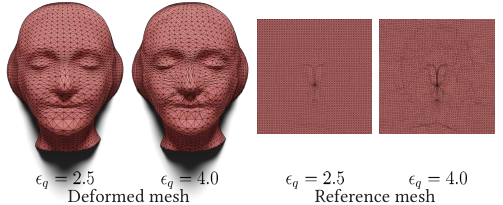}
    \caption{\textbf{Ablation on the mesh quality tolerance $\epsilon_q$} We compare two settings of $\epsilon_q$ on a cloth draped over a face (Figure~10). \textbf{Left:} the deformed mesh. \textbf{Right:} the corresponding adapted reference mesh. At $\epsilon_q = 2.5$ (our default, used uniformly across all examples in the paper), the adapted mesh faithfully captures the underlying facial geometry while maintaining clean, near-isotropic triangulation in both configurations. Raising the tolerance to $\epsilon_q = 4.0$ relaxes the quality constraint and admits more aggressive anisotropic adaptation: triangles elongate along feature directions in the deformed mesh, and the resulting vertex clustering produces visible sliver triangles in the reference mesh. The larger tolerance therefore yields finer geometric tracking at the cost of element quality, motivating our conservative default.}
    \label{fig:ablation_study}
\end{figure}


\begin{figure}[!h]
    \centering
    \includegraphics[width=1.0\linewidth]{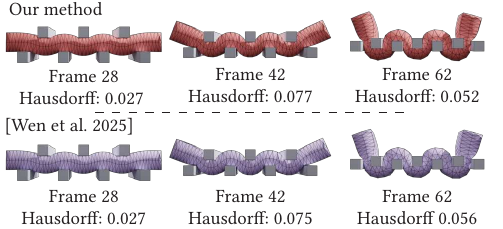}
    \caption{\textbf{Performance comparison on the 3D masticator benchmark.} We compare our method (top) against the original r-ITR framework~\cite{Wen:2025:ORI} (bottom) on a volumetric simulation. Visuals: as shown in the three representative frames, our results are visually indistinguishable from the baseline. Performance: our method achieves a $6\times$ speedup for the full animation sequence (518.98s vs. 3103.21s). Accuracy: this efficiency gain comes with negligible accuracy loss. Compared to a high-resolution fixed mesh simulation (71K vertices), our average Hausdorff distance increases by only 1\% (0.04658 vs. 0.04611), demonstrating that our method accelerates convergence without compromising physical fidelity.}
    \label{fig:masticator}
\end{figure}

\begin{table}[!h]
\begin{tabular}{|l|l|l|l|l|l|l|}
\hline
Example & \#V & \#F & $E(Pa)$ & $\nu$ & $\rho (kg/m^3)$  & $h$(mm) \\ \hline
Fig.1 (left)       & 14,413    &  27,834   &  6e5   &   0.3    &  500      & 0.2       \\ \hline
Fig.1 (right)       &   31,367  & 62,071    &  8e5  &    0.24   &  472.6    & 0.3       \\ \hline
Fig.3        &   221     &  400      &  2e5  &    0.3    &  1,000    & 0.2       \\ \hline
Fig.7       &  3,281    &  6,400    &  2e4  &    0.3    &  1,000    & 0.1       \\ \hline
Fig.9     &   1,099   & 1,968     &  1e4  &    0.1    &  1,000    & 0.1       \\ \hline
Fig.10        &  20,201   & 40,000    &  6e3  &    0.3    &  1,000    & 0.01      \\ \hline
\end{tabular}
\caption{\textbf{Material parameters for all examples.} \#V: vertex count, \#F: triangle count, $E$: Young's modulus, $\nu$: Poisson's ratio, $\rho$: density, $h$: thickness. All examples use neo-Hookean elasticity, except for the teaser (Figure~1) which uses StVK.}
\label{table:material}
\end{table}

\bibliographystyle{ACM-Reference-Format}
\bibliography{references}